\documentclass[aps,prl,twocolumn,superscriptaddress,amsmath,amssymb,floatfix]{revtex4-2}
\usepackage{graphicx}
\usepackage{bm}
\usepackage{xcolor}
\usepackage[colorlinks=true,linkcolor=black,citecolor=black,urlcolor=black]{hyperref}

\newcommand{\msun}{M_\odot}
\newcommand{\GV}{G_V}
\newcommand{\GVmin}{G_V^{\rm min}}
\newcommand{\DMB}{\Delta M_B}
\newcommand{\muc}{\mu_{c,\max}}
\newcommand{\rev}[1]{#1}
\newcommand{\revb}[1]{#1}
\newcommand{\revc}[1]{#1}

\begin{document}

\title{Neutron stars nearly close the window for muon-induced baryon-number violation}

\author{Bharat Kumar}
\email{kumarbh@nitrkl.ac.in}
\affiliation{Department of Physics and Astronomy, National Institute of Technology, Rourkela 769008, India}

\date{\today}

\begin{abstract}
Baryon-number violation that needs a muon in the initial state can evade proton-decay searches and be observed in muon capture at Mu2e and COMET. Neutron stars contain muons, so the same interaction converts their protons into dark baryons, but the rate of this conversion in dense matter and its consequences have not been computed. In this paper we compute the conversion rate, the heat it deposits, and the evolution of a star under fast conversion. The coldest known neutron stars exclude conversion rates far below laboratory reach for every dark-fermion mass that capture in aluminium can produce, and faster conversion burns stars down or, unless conserved dark baryons repel each other strongly enough, collapses them. The repulsion they need follows a scaling law in the baryonic mass headroom of the nucleonic equation of state, which applies to any process that puts dark baryons in equilibrium with neutrons.
\end{abstract}

\maketitle

\section{Introduction}

Baryon-number violation (BNV) is required to generate the baryon asymmetry of the Universe~\cite{Sakharov1967}, while the Standard Model conserves baryon number to all orders in perturbation theory. Searches for proton decay~\cite{SuperK2020} and for invisible neutron decay in nuclei~\cite{KamLAND2006} constrain BNV tightly when the initial state contains only nucleons, but they are far less sensitive to BNV that needs a muon in the initial state, because a free proton can then decay only through a virtual muon, and a suitable dark mass spectrum makes this decay extremely slow~\cite{Fox2024}. Fox, Hostert, Menzo, Pospelov and Zupan proposed a model of this kind~\cite{Fox2024}, in which a muon bound in an atom is captured by a proton into two dark fermions,
\begin{equation}
\mu^- + p \;\to\; \chi_0 + \chi_2, \qquad \chi_2 \to \chi_1 + A',
\label{eq:reaction}
\end{equation}
where $\chi_0$ is a light neutral fermion and the dark photon $A'$ decays to an $e^+e^-$ pair. Either $\chi_0$ or the pair $\chi_1,\chi_2$ can carry the baryon number of the proton, so the violation of baryon number is only apparent~\cite{Fox2024}. The Mu2e~\cite{Mu2e,Mu2eII} and COMET~\cite{COMETTDR,COMET} experiments, built to search for coherent muon-to-electron conversion in nuclei~\cite{Calibbi2018}, can detect the $e^+e^-$ pair in their aluminium stopping targets~\cite{Fox2024}, and the benchmark models of Ref.~\cite{Fox2024} predict rates within their reach while a suitable choice of the dark masses avoids the limits from proton decay and nuclear stability.

The same reaction runs inside neutron stars, because beta equilibrium creates muons wherever the electron chemical potential exceeds the muon mass. Their abundance depends on the nuclear symmetry energy~\cite{ZhangLi2020}, and in our model stars the core holds several muons per hundred baryons~\cite{SM}. Stellar muons already constrain the axion-muon coupling through supernovae~\cite{Bollig2020} and muonic fifth forces through neutron stars~\cite{Fiorillo2026}. Reaction~\eqref{eq:reaction} runs in every neutron star whose central neutron chemical potential exceeds the $\chi_2$ mass, and Fox \emph{et al.} pointed out its two consequences~\cite{Fox2024}. A slow conversion heats the star, which the coldest neutron stars can test~\cite{McKeen2021,Guillot2019}, whereas a fast conversion drives the star to a new equilibrium with dark baryons, which must still support the observed neutron-star masses, as studies~\cite{McKeen2018,Baym2018,Motta2018} of the dark neutron decay proposed in Ref.~\cite{FornalGrinstein2018} have shown.

Neither consequence has been quantified, and the available estimate of the stellar rate treats the proton as free and leaves unevaluated the phase-space average, which is small near threshold~\cite{Fox2024}, whereas in a star the proton is a relativistic quasiparticle with an effective mass and a vector potential. The equilibrium requirement depends, in turn, on the nucleonic equation of state (EoS), which is not known at the densities of the core, so any bound on the dark force inherits the uncertainty of dense-matter physics.

In this paper, we compute the in-medium conversion rate and the heat it deposits throughout the star, follow fast conversion to its end, and compute exotic capture in aluminium with its nuclear structure. We find that heating excludes slow conversion far below the laboratory reach, and that the repulsion a star needs to survive fast conversion follows a scaling law in the baryonic mass headroom of the nucleonic EoS.

\section{Method}

We describe cold, beta-equilibrated $npe\mu$ matter with twenty relativistic mean-field (RMF) parameter sets. Nineteen are the sets of the compilation of Sun, Bhattiprolu and Lattimer~\cite{Sun2024} whose beta-equilibrated stars, computed with our crust, reach $2.01\,\msun$, the lower edge of the 68\% credible interval for the mass of PSR J0740+6620~\cite{Fonseca2021}, and satisfy the bound $\tilde\Lambda\le720$ on the tidal deformability from GW170817~\cite{Abbott2019}; the twentieth is GM3~\cite{GM1991}. We use natural units, $\hbar=c=k_B=1$, and the Supplemental Material gives the details and validations of the calculation~\cite{SM}. It also shows that the Skyrme force SLy4~\cite{Chabanat1998}, whose cores are denser, gives a heating bound an order of magnitude stronger, so the RMF results are conservative.

In each model star we evaluated the conversion rate per muon for the operator $\mathcal{L}=(G/\sqrt2)(\bar p\chi_2)(\bar\mu\chi_0)$ of Ref.~\cite{Fox2024} by Monte Carlo integration over the degenerate proton and muon Fermi seas, with exact energy-momentum conservation. We used vacuum spin sums in which the proton carries its RMF Dirac effective mass and an energy shifted by the vector potential, as in relativistic calculations of Urca rates~\revc{\cite{Alford2021,Alford2024}}. The benchmark masses place the reaction close to threshold, where the final pair carries little momentum, and this small two-body phase space, absent from the estimate $\Gamma\simeq G^2m_\mu m_p n_p$ of Ref.~\cite{Fox2024}, makes our rate orders of magnitude smaller than that estimate except in the heaviest stars. We integrated the local rate over the volume of each star with the gravitational redshift, and the conversion runs mainly in the inner core, where the reaction is above threshold.

Each conversion heats the star in two ways: the dark photon gives its energy to thermal electrons and photons, and the proton and muon removed from below their Fermi surfaces leave holes that are refilled~\cite{Fox2024}. We do not count the kinetic energy of $\chi_1$, which exchanges energy with matter only at second order in the dark-photon coupling~\cite{SM}, so the heat is a lower bound. The luminosity at infinity is
\begin{equation}
L_\infty=\int dV\, e^{2\Phi}\,\Gamma\, n_\mu\, \langle Q\rangle ,
\label{eq:heat}
\end{equation}
where $dV$ is the proper volume element, $e^{\Phi}$ is the lapse function, $\Gamma n_\mu$ is the local conversion rate per volume and $\langle Q\rangle$ is the mean heat per conversion. In benchmark~I the vector coupling of the $A'$ arises from maximally mixed Majorana states~\cite{Fox2024}, so $\chi_1$ is its own antiparticle, dark baryon number is not conserved, and $\chi_1$ pairs annihilate, $\chi_1\chi_1\to A'A'$, through $\chi_2$ exchange. Annihilation is then much faster than production~\cite{SM}, so the energy of $\chi_1$, including its rest mass, also becomes heat, and $\langle Q\rangle$ rises several-fold.

The dark photon deposits its energy \revc{inside the star}. It cannot decay in the core, because the electron Fermi energy exceeds the $A'$ energy and Pauli blocks the final electron, so it scatters instead on the degenerate electrons through its kinetic mixing $\varepsilon$, $A'e\to\gamma e$, after which the photon thermalizes at once. \revc{The vector current of benchmark~I connects fermions of different mass and emits almost only longitudinal $A'$, which couple to the electron current only through their mass, whereas the dipole of benchmark~II emits transverse $A'$. With Pauli blocking and the in-medium mixing~\cite{An2013} of each polarization in a degenerate plasma~\cite{BraatenSegel1993}, the benchmark mixing $\varepsilon=10^{-4}$ deposits the energy within a metre in benchmark~I and within about a centimetre in benchmark~II, and the $A'$ leaves the star only for $\varepsilon\lesssim6\times10^{-7}$ and $10^{-7}$, respectively~\cite{SM}. Its decay length grows as $\varepsilon^{-2}$, from a few millimetres at the benchmark mixing~\cite{Fox2024} to tens of metres or more wherever it escapes, so there most $A'$ decay outside the stopping targets of Mu2e and COMET.}

Heat is the signature of slow conversion, whereas fast conversion ends in a new equilibrium if $\chi_1$, which we then call $X$, carries a conserved dark baryon number, because conversion then fills a dark Fermi sea. The reaction has no inverse, since $\chi_0$ leaves the star, so conversion stops when the dark states it feeds are Pauli blocked, which for thermalized dark baryons happens at
\begin{equation}
\mu_X \;=\; \mu_p + \mu_\mu \;=\; \mu_n \;\equiv\; \mu_B ,
\label{eq:equil}
\end{equation}
up to the small $\chi_0$ mass. The redshifted chemical potential of each conserved species is uniform in a static star, so Eq.~\eqref{eq:equil} holds at every radius. The dark baryon has chemical potential $\mu_X=\sqrt{k_X^2+m_X^2}+\GV n_X$ and pressure $P_X=P_{\rm FG}(k_X)+\tfrac12\GV n_X^2$, where $P_{\rm FG}$ is the Fermi-gas pressure and $\GV=g_D^2/m_V^2$ is the strength of a repulsive vector interaction among dark baryons~\cite{McKeen2018,Cline2018}. In the benchmark models the dark photon couples $\chi_1$ to $\chi_2$ off diagonally and generates no mean field, so $\GV=0$. The repulsion that Ref.~\cite{Fox2024} proposes to restore the pressure therefore requires an additional, diagonal dark charge, \revc{The alternative it mentions, a cascade of $\chi_1$ into several lighter dark fermions, divides the baryon chemical potential among them and can leave stars of about $2\,\msun$ without repulsion~\cite{Strumia2022,Darini2023}; we do not treat it.} To test how the equilibrium depends on the EoS, we also continued each RMF set above $1.5\,n_0$ with a constant squared sound speed $c_s^2=0.5$ or $1$, which gives sixty EoSs with maximum masses from $2.0$ to $3.4\,\msun$.

\section{Results and discussion}

The heating bound rests on PSR J2144$-$3933, an isolated pulsar with a characteristic age of about $3\times10^8$~yr and the coldest known neutron star, whose thermal emission has never been detected~\cite{Guillot2019}. The recalibrated upper limit on its redshifted surface temperature is $3\times10^4$~K~\cite{Rodriguez2025}. In such an old, cold star neutrino emission is negligible~\cite{YakovlevPethick2004}, so internal heat must leave through the surface~\cite{Rodriguez2025}, and Eq.~\eqref{eq:heat} cannot exceed the blackbody luminosity at that temperature, about $10^{27}$~erg~s$^{-1}$. Other heat sources, such as rotochemical heating~\cite{Reisenegger1995} and vortex creep~\cite{Guillot2019,Rodriguez2025}, only add to the luminosity and make the bound conservative. The same argument, applied to dark-matter capture~\cite{Baryakhtar2017}, has constrained mirror neutrons~\cite{McKeen2021} and dark-matter-catalyzed baryon destruction~\cite{Ema2025}, while binary-pulsar period changes limit baryon loss independently~\cite{Berryman2024}. As a second star we used PSR J0437$-$4715, a millisecond pulsar of $1.42\,\msun$~\cite{Reardon2024} whose white-dwarf companion is about $6$~Gyr old~\cite{Durant2012}, for which a hydrogen-atmosphere fit to the ultraviolet and soft X-ray spectra gives a redshifted surface temperature of $2.4\times10^5$~K~\cite{Rodriguez2025}.

\begin{figure}[t]
\includegraphics[width=\columnwidth]{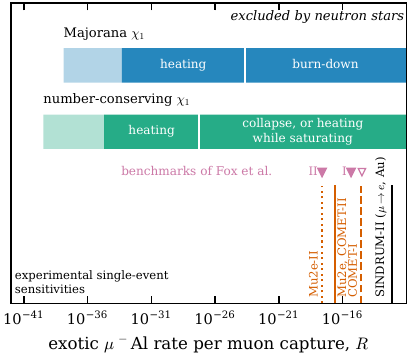}
\caption{\rev{Exotic $\mu^-$Al capture rate per ordinary muon capture, $R$, excluded by neutron stars at the benchmark masses, for a Majorana $\chi_1$ (upper bar) and a number-conserving $\chi_1$ (lower bar). Heating: Eq.~\eqref{eq:heat} exceeds the upper limit on the thermal emission of PSR J2144$-$3933~\cite{Rodriguez2025}. Burn-down: a $1.4\,\msun$ star falls below $1.174\,\msun$ within $1$~Gyr. Collapse: a $1.4\,\msun$ star with $\GV<\GVmin$ collapses within $1$~Gyr, or, for a collisionless dark gas, still outshines the heating limit~\cite{SM}. Dark bars give the most conservative case over twenty RMF models and, for heating, over stellar masses from $1.2\,\msun$ to the maximum mass; light parts show the spread. Vertical lines give the single-event sensitivities of COMET Phase~I, of Mu2e and COMET Phase~II ($2$--$3\times10^{-17}$) and of Mu2e-II~\cite{Mu2e,Mu2eII,COMETTDR,COMET}, \revb{and, for orientation, the SINDRUM-II limit on coherent $\mu\to e$ conversion in gold~\cite{SINDRUM2006}, which does not apply directly to the $e^+e^-$ signal}. The upper bar uses the aluminium rate of benchmark~I, including the off-shell continuum, and the lower bar that of benchmark~II. Filled triangles: the two benchmarks of Ref.~\cite{Fox2024} with our rates; open triangle: the estimate of Ref.~\cite{Fox2024} for both.}}
\label{fig:reach}
\end{figure}

In Fig.~\ref{fig:reach} we show the exotic capture rates in aluminium, the observable of Mu2e and COMET, that neutron stars exclude at the benchmark masses. The stellar and laboratory rates both scale as $G^2$, and our calculation of capture in aluminium fixes the map between them for each benchmark. We see that heating excludes rates down to $5\times10^{-34}$ for the most conservative model and stellar mass, more than sixteen orders of magnitude below the single-event sensitivity of Mu2e. Heavier stars give stronger bounds, because a larger part of their cores lies above threshold, and the bound depends on neither the dark repulsion nor the thermal state of the dark baryons.

\begin{figure}[t]
\includegraphics[width=\columnwidth]{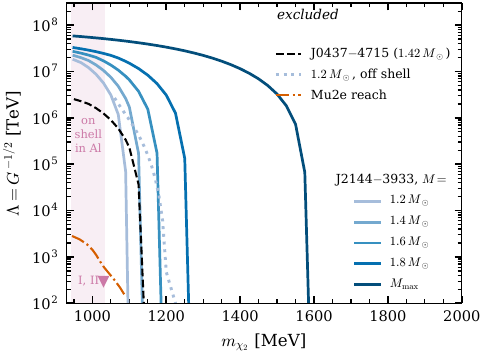}
\caption{\rev{Lower bound on the operator scale $\Lambda=G^{-1/2}$ from internal heating, against the $\chi_2$ mass, with $m_{\chi_1}=m_{\chi_2}-130$~MeV and $m_{A'}=20$~MeV as in the benchmarks. Solid lines: PSR J2144$-$3933~\cite{Rodriguez2025} for stellar masses of $1.2$, $1.4$, $1.6$ and $1.8\,\msun$ and the maximum mass of each EoS ($2.0$--$2.3\,\msun$), from light to dark. Dashed line: PSR J0437$-$4715 with its measured mass~\cite{Reardon2024} and surface temperature~\cite{Rodriguez2025}, taken two quoted uncertainties above the central value. Each line gives the most conservative of twenty RMF models and ends where $m_{\chi_2}$ reaches the smallest central neutron chemical potential among them. The shaded band marks the $\chi_2$ masses that muon capture in aluminium produces on shell for this splitting. The triangle marks benchmarks I and II.} \rev{Dotted line: $1.2\,\msun$ bound with the off-shell channel. Dash-dotted line: Mu2e reach, $R=2.5\times10^{-17}$, from our three-body rate in aluminium for $\alpha_D=10^{-3}$, assuming full acceptance~\cite{SM}.}}
\label{fig:m2}
\end{figure}

In Fig.~\ref{fig:m2} we show how the heating bound on the operator scale $\Lambda=G^{-1/2}$ depends on the $\chi_2$ mass and on the mass of the star. We see that at the benchmark mass the bound lies more than three orders of magnitude above the Mu2e reach for every stellar mass, and even for the warmer PSR J0437$-$4715. Each curve ends where $m_{\chi_2}$ equals the central neutron chemical potential, so a heavier star, with its denser centre, bounds heavier $\chi_2$, up to $1.6$--$2.0$~GeV at the maximum mass. Since the mass of PSR J2144$-$3933 is unknown, we quote the weakest curve, for $1.2\,\msun$, as our bound.

The laboratory signal, in contrast, is confined to a narrow range of $\chi_2$ masses, shaded in Fig.~\ref{fig:m2}. Capture in aluminium produces an on-shell $\chi_2$ only if $m_{\chi_0}+m_{\chi_2}<m_\mu+m_p-S_p-E_b\approx1035$~MeV, where $S_p=8.27$~MeV is the proton separation energy of $^{27}$Al and $E_b$ is the muon binding energy, while proton stability requires $m_{\chi_1}+m_{A'}>m_p-m_\mu$~\cite{Fox2024}, so that $m_{\chi_2}\gtrsim943$~MeV for the benchmark splitting. In all our EoSs every neutron star heavier than $1.174\,\msun$, the mass of the companion of PSR J0453+1559, very likely a neutron star~\cite{Martinez2015}, has a central neutron chemical potential above this range. The heating bound therefore covers the whole range for any mass of PSR J2144$-$3933 above that value.

Within this range the map between $G$ and $R$ requires the capture rate in the nucleus, which Ref.~\cite{Fox2024} estimated by scaling ordinary muon capture with a ratio of Dalitz areas. We computed it instead in the impulse approximation, with the bound muon, Woods-Saxon proton orbitals that reproduce the proton separation energy of $^{27}$Al, and a $^{26}$Mg nucleus left with a hole in the orbital of the captured proton~\cite{SM}. The same code gives capture rates consistent with the measured rate in muonic hydrogen~\cite{MuCap2013} and, within the factor expected from the neglect of Pauli blocking, with that in aluminium~\cite{Suzuki1987,SM}. Near threshold the on-shell rate falls far below the estimate, because only the outermost $1d_{5/2}$ protons can produce $\chi_2$ and the scalar matrix element vanishes with the energy of the massless $\chi_0$. At the benchmark mass we find $R=2.5\times10^{-18}$, ten times below the Mu2e sensitivity, so the narrow $\chi_2$ of benchmark~II gives no signal at Mu2e. The broad $\chi_2$ of benchmark~I, in contrast, is produced mostly off shell, $\mu^-p\to\chi_0\chi_1A'$ through a virtual $\chi_2$, in a channel open to every proton shell, which gives $R\approx5\times10^{-16}$, seven times below the estimate. Spreading of the proton-hole strength over excited states of $^{26}$Mg makes these rates uncertain by a factor of a few~\cite{SM}.

The off-shell channel also extends the laboratory signal beyond $1035$~MeV, and it runs in stars as well, where we computed it with the full $\chi_2$ propagator~\cite{SM}. The laboratory channel closes at $m_{\chi_2}\approx1145$~MeV, and up to that point the off-shell bound from a $1.2\,\msun$ star stays more than three orders of magnitude above the Mu2e reach (Fig.~\ref{fig:m2}).

All of these bounds assume that conversion continues at the present age of the pulsar~\cite{McKeen2021}, and at faster rates a star exhausts the conversion sooner. If $\chi_1$ annihilates, the dark sector never fills, and conversion continues until the central chemical potential falls below the $\chi_2$ mass, so the star burns down below the mass of the lightest known neutron star. At the benchmark coupling a $1.4\,\msun$ star falls below $1.174\,\msun$ within one to five years, and recycled pulsars with measured masses near $2\,\msun$, such as PSR J0740+6620~\cite{Fonseca2021}, therefore exclude the right part of the upper bar in Fig.~\ref{fig:reach}. At slower rates PSR J2144$-$3933, younger than the $1$~Gyr that defines that bar, is still burning, so the two arguments overlap, and a Majorana $\chi_1$ is excluded at all rates above the heating bound, whatever the dark repulsion.

A conserved $\chi_1$ changes the end point, because fast conversion then ends in the dark equilibrium of Eq.~\eqref{eq:equil}. Without repulsion no EoS supports a $1.174\,\msun$ star, and the two-species end state of the benchmark models, with $\chi_2$ and $\chi_1$ seas in equilibrium, supports only about $0.6\,\msun$~\cite{SM}, somewhat below the $0.7$--$0.8\,\msun$ found for a single non-interacting dark baryon at the neutron mass~\cite{McKeen2018,Baym2018,Motta2018,Husain2022}. If the dark baryons thermalize, a star collapses once a few percent of its baryons have converted, which at the benchmark coupling takes weeks to months, and a star that has not yet collapsed is still converting and still heating.

Collapse within weeks requires the dark baryons to thermalize, which is not guaranteed~\cite{Husain2022}: the benchmark dark sector has no elastic self-interaction at tree level, and $\chi_1$ scatters on protons and electrons only at second order, through a virtual $\chi_2$ and two dark photons. We computed this energy loss in the degenerate plasma~\cite{SM} and find that with the vector coupling of benchmark~I a conserved $\chi_1$ would cool within a year, whereas with the transition dipole of benchmark~II, the number-conserving case, it keeps its birth energy for about $10^9$~yr. We therefore solved for the equilibrium of a star that contains such a collisionless dark gas, with the occupation of each orbit frozen in its radial action and angular momentum~\cite{SM}. The gas does not collect in a dense centre, so the star survives until about a fifth of its baryons have converted. Since conversion slows as the orbits it feeds fill up, collapse at the benchmark coupling takes centuries, and during this time the star outshines the limit from PSR J2144$-$3933 by many orders of magnitude, so the number-conserving bar of Fig.~\ref{fig:reach} stays excluded whether or not the dark baryons cool.

A conserved dark baryon thus survives only if its repulsion is strong enough. Conversion conserves baryon number, so a star evolves at fixed baryonic mass $M_B$, as noted for dark neutron decay~\cite{Motta2018,Husain2022}, and survives only if the converted EoS supports the same baryonic mass. We define $\GVmin$ by the condition that the converted EoS supports the baryonic mass of a $2\,\msun$ nucleonic star, and we measure the room that the nucleonic EoS leaves by the baryonic headroom $\DMB = M_{B,\max}^{(0)} - M_B(2\,\msun)$, the counterpart of the gravitational headroom that sets neutron-star limits on dark matter carried along with the baryons~\cite{Kalita2026}. Once the repulsion dominates the dark Fermi energy, the dark density is limited to $n_X\simeq(\mu_B-m_X)/\GV$, so the baryons that the dark sea absorbs scale as $1/\GV$ and must stay below the headroom, which suggests the scaling form
\begin{equation}
\GVmin \;=\; A(m_X)\,\left(\frac{\DMB}{\msun}\right)^{-a}\left(\frac{\muc-m_X}{100~{\rm MeV}}\right)^{b},
\label{eq:law}
\end{equation}
where $\muc$ is the central baryon chemical potential of the nucleonic maximum-mass star and $A$ is in fm$^2$.

For dark-baryon masses from $875$ to $1250$~MeV, the sixty EoSs, which span a factor of a hundred in headroom and include sound speeds up to the causal limit, follow Eq.~\eqref{eq:law} with an exponent $a$ close to one and a root-mean-square scatter of two to eight percent; the Supplemental Material gives the fit and its domain of validity~\cite{SM}. Without refitting, Eq.~\eqref{eq:law} also predicts $\GVmin$ for Skyrme forces~\cite{Sun2024}, for hybrid EoSs with a first-order transition to quark matter of constant sound speed~\cite{AlfordHan2013}, and for EoSs whose maximum-mass cores hold up to about forty percent hyperons with SU(3) vector couplings that keep them above $2\,\msun$~\cite{Weissenborn2012}, to eight percent root-mean-square and seventeen percent at worst. A bosonic dark baryon follows a law of the same form with up to a third more repulsion, because a condensate has no Fermi pressure, and the hyperonic EoSs obey it too~\cite{SM}.

Earlier studies of repulsive dark baryons~\cite{McKeen2018,Cline2018,Grinstein2019,Husain2022} derived the required repulsion one EoS at a time, and a recent scan over twenty-one RMF EoSs~\cite{RafieiKarkevandi2026} found that, beyond the weakest repulsions, the excluded region depends on the EoS. Equation~\eqref{eq:law} traces this dependence to the headroom and turns a statement about the EoS into a statement about the dark sector: a bound on the headroom of the true EoS from observations translates directly into a lower bound on the dark repulsion. An EoS with a maximum mass only slightly above $2\,\msun$ requires a repulsion of order $10^2$~fm$^2$, which for a dark coupling of order one points to a vector mediator lighter than about $20$~MeV.

\section{Conclusion}

We have shown that neutron stars exclude muon-induced baryon-number violation far beyond the reach of laboratory experiments. The heat deposited in the coldest known neutron stars excludes conversion rates far below the Mu2e sensitivity for every $\chi_2$ mass that capture in aluminium can produce, on or off shell, whatever the dark repulsion. Faster conversion burns the star down if $\chi_1$ annihilates, as in benchmark~I, and if the dark baryons are conserved and repel each other too weakly, it keeps the star hot until it collapses, whether or not the dark baryons cool. Capture in aluminium is itself slower near threshold than estimated, so the number-conserving benchmark lies below the Mu2e sensitivity even without the stellar bound.

An on-shell signal at Mu2e or COMET from this operator therefore requires a conserved dark baryon that converts the star quickly and survives through a repulsion above the scaling law of Eq.~\eqref{eq:law}, or a dark sector in which $\chi_1$ cascades into several lighter fermions, which Ref.~\cite{Fox2024} proposes and which \revc{lets heavy stars survive without repulsion~\cite{Strumia2022,Darini2023,SM}}; we have not computed the stars of this second case. The same requirement applies to any process that puts a dark baryon in equilibrium with neutrons, such as the dark decay of the neutron~\cite{FornalGrinstein2018}. On the laboratory side the largest remaining uncertainty is the distribution of proton-hole strength in $^{26}$Mg near threshold, which a shell-model calculation with measured spectroscopic factors would fix.

\clearpage
\onecolumngrid
\setcounter{section}{0}\setcounter{equation}{0}\setcounter{figure}{0}\setcounter{table}{0}
\renewcommand{\thefigure}{S\arabic{figure}}
\renewcommand{\thetable}{S\arabic{table}}
\renewcommand{\theequation}{S\arabic{equation}}
\begin{center}
{\large\bfseries Supplemental Material: Neutron stars nearly close the window for muon-induced baryon-number violation}\\[4pt]
B. Kumar\\
{\small Department of Physics and Astronomy, National Institute of Technology, Rourkela 769008, India}
\end{center}

\section{Equations of state and stellar structure}

The twenty nucleonic EoSs are the nineteen RMF sets of Ref.~\cite{Sun2024} whose beta-equilibrated stars, computed with our SLy crust, satisfy the mass bound from PSR J0740+6620~\cite{Fonseca2021} and the tidal-deformability bound from GW170817~\cite{Abbott2019} quoted in the Letter, together with GM3~\cite{GM1991}. The nineteen sets are FSU2R, GL2, HA, HC, NLM, NLM2, NLM6, NL$\rho$, PCSB2, QMC-RMF2, QMC-RMF3, QMC-RMF4, RMF415, RMF418, RMF419, RMF421, RMF422, RMF423 and TW-99, and we select them with our own stars, because the neutron-star properties tabulated in Ref.~\cite{Sun2024} refer to pure-neutron-matter cores. Each set is joined to the SLy crust with a single nucleon mass. The selection combines the lower edge of a $68\%$ interval with the upper end of a $90\%$ interval and does not use the NICER radius of PSR J0740+6620, below which TW-99 and GL2 fall~\cite{Kalita2026}; removing sets can only strengthen the most conservative heating bound, and the scaling law holds set by set, so neither result depends on this choice. The extended EoSs replace each set above $1.5\,n_0$ by the constant-sound-speed form $P=P_t+c_s^2(\varepsilon-\varepsilon_t)$ of Ref.~\cite{AlfordHan2013}, without a jump, with $c_s^2=0.5$ or $1$, and the baryon density follows from $d\ln n=d\varepsilon/(\varepsilon+P)$.

We solved
\begin{align}
\frac{dP}{dr}&=-\frac{(\varepsilon+P)(m+4\pi r^3P)}{r(r-2m)},\qquad \frac{dm}{dr}=4\pi r^2\varepsilon,\\
\frac{dM_B}{dr}&=\frac{4\pi r^2 m_N n_B}{\sqrt{1-2m/r}},\qquad \frac{d\Phi}{dr}=\frac{m+4\pi r^3P}{r(r-2m)},
\end{align}
with $e^{2\Phi(R)}=1-2M/R$. The solver passes the uniform-density test, is converged in the step size, reproduces the maximum mass and radius of an independent RMF code for FSU2R, and satisfies $dP/d\mu_B=n_B$ for the mixed EoSs. In every set the central neutron chemical potential of a $1.174\,\msun$ star lies above the on-shell threshold for capture in aluminium, and that of a $1.2\,\msun$ star lies above the benchmark $\chi_2$ mass. For a non-interacting dark baryon at the neutron mass the maximum mass is about $0.7\,\msun$, as Refs.~\cite{McKeen2018,Baym2018,Motta2018} find, and for heavier dark baryons it stays close to the values of Ref.~\cite{McKeen2018}. The two-fluid TOV equations~\cite{Husain2022}, with the dark baryon coupled to the nucleons only by gravity and equal redshifted chemical potentials, agree with the single-fluid solution.

\section{Conversion rate and heat}\label{sec:rate}

The spin-summed squared matrix element is $\sum|\mathcal M|^2=8G^2(p_2\cdot p_p^*+m_2M^*)(p_0\cdot p_\mu)$, where $p_p^*=(E^*,\bm p)$ with $E^*=\sqrt{p^2+M^{*2}}$, and the proton energy is $E^*+V_p$. The rate per volume is
\begin{equation}
\Gamma n_\mu=\int\frac{d^3p_\mu}{(2\pi)^3}\frac{d^3p_p}{(2\pi)^3}\frac{1}{2E_\mu\,2E^*}\int d\Phi_2\,\sum|\mathcal M|^2 ,
\end{equation}
over the muon and proton Fermi spheres, with the dark final states empty. We use vacuum spin sums in which the proton carries its in-medium Dirac mass and an energy shifted by the vector potential, \revc{the convention of the relativistic Urca rates of Ref.~\cite{Alford2021}, also used in Ref.~\cite{Alford2024}, in which the unshifted energy $E^*=\sqrt{p^2+M^{*2}}$ enters the phase-space normalization and the Dirac traces while energy conservation involves $E^*$ plus the vector potential.} We sampled both momenta uniformly in their Fermi spheres, evaluated the two-body phase space exactly in the centre-of-mass frame of each pair, and boosted to the star frame, and in vacuum this integration reproduces the analytic two-body rate $\Gamma=G^2(E_2+m_2)k^2n_p/(8\pi\sqrt s)$, where $k$ is the momentum of the final pair. The dimensional estimate $\Gamma\simeq G^2m_\mu m_pn_p$ of Ref.~\cite{Fox2024} exceeds this vacuum rate by more than three orders of magnitude at the benchmark masses, because the two-body phase space is small near threshold, and in the core the Fermi motion enlarges the phase space and reduces the gap, most in the heaviest stars. \revc{Our matrix element follows from the scalar operator that Ref.~\cite{Fox2024} uses for its benchmarks. The ultraviolet completion of that reference has a chiral structure, which changes the capture rates by factors of order one~\cite{Fox2024}, far less than the many orders of magnitude between the heating bound and the laboratory reach.}

The heat that each conversion deposits is the lab-frame energy of the $A'$, averaged over the isotropic $\chi_2$ decay, $E_{A'}=(E_2/m_2)(m_2^2-m_1^2+m_{A'}^2)/(2m_2)$, plus the hole energy $(\mu_p+\mu_\mu)-(E_p+E_\mu)$, and for a Majorana $\chi_1$ also the energy of $\chi_1$, which raises the luminosity at the benchmark masses by a factor of five to eight. Removing a proton and a muon also drives the matter out of beta equilibrium, and the weak reactions that restore it release further energy, partly as heat and partly in neutrinos; we do not count this energy, nor the recoil energy of $\chi_1$, so the heat is a lower bound.

\section{The dark photon in the core}

In vacuum the benchmark $A'$ decays within a few millimetres~\cite{Fox2024}, but in the core the electron from $A'\to e^+e^-$ must lie above the electron Fermi energy, which at the centres of our stars is $\mu_e=176$--$219$~MeV and exceeds the $A'$ energy, so the decay is blocked, and only a few percent of the decays remain open even at half the central density. \revc{The $A'$ is instead absorbed by scattering on the degenerate electrons, $A'e\to\gamma e$, after which the photon thermalizes at once, and the rate depends on its polarization. The vector current $\bar\chi_2\gamma^\mu\chi_1$ of benchmark~I connects fermions of different mass, so its divergence is proportional to $m_2-m_1$ and it emits longitudinal $A'$ with a strength enhanced by $(m_2-m_1)^2/m_{A'}^2$; we find that more than $99.9\%$ of the $A'$ from $\chi_2\to\chi_1A'$ are longitudinal. The transition dipole of benchmark~II couples to the field strength and emits only transverse $A'$. The in-medium mixing of each polarization is $\varepsilon_{T,L}=\varepsilon m_{A'}^2/|m_{A'}^2-\Pi_{T,L}|$~\cite{An2013}, with the polarization functions of a degenerate ultrarelativistic electron gas~\cite{BraatenSegel1993}, where the longitudinal function of Ref.~\cite{An2013} is that of Ref.~\cite{BraatenSegel1993} multiplied by $(\omega^2-k^2)/k^2$. The transverse mixing is enhanced by up to about a factor of two, whereas the longitudinal mixing stays within five percent of $\varepsilon$. We evaluated the scattering rate of each polarization by Monte Carlo integration over the electron Fermi sea, with explicit Dirac traces for a massive vector of the given polarization and Pauli blocking of the final electron, which removes most of the free rate; for a massless transverse photon the traces reproduce the Klein-Nishina cross section. A longitudinal $A'$ couples to the conserved electron current only through its mass, and its mean free path is fifteen to forty times that of a transverse $A'$ of the same energy. In $1.2$ and $1.4\,\msun$ stars the mean free path is $(2$--$11)\times10^{-9}\,{\rm cm}/\varepsilon^2$ for transverse and $(0.6$--$3.8)\times10^{-7}\,{\rm cm}/\varepsilon^2$ for longitudinal $A'$, so at $\varepsilon=10^{-4}$ the energy is deposited within about a centimetre in benchmark~II and within a metre in benchmark~I, and it stays in the star for $\varepsilon\gtrsim10^{-7}$ and $\varepsilon\gtrsim6\times10^{-7}$, respectively. Muons and protons, whose contributions to the polarization functions scale with their Fermi velocities~\cite{BraatenSegel1993}, are left out; for the transverse mode they bring $\Pi_T$ closer to $m_{A'}^2$ and raise $\varepsilon_T$, so the electron-only value overestimates the mean free path. For smaller mixing the bound rests on the hole energy alone and is correspondingly weaker, and part of the hole energy leaves in Urca neutrinos. The vacuum decay length grows as $\varepsilon^{-2}$ from a few millimetres at $\varepsilon=10^{-4}$~\cite{Fox2024}, so where the $A'$ escapes the star it decays tens of metres or more from the point of capture, and most decays fall outside the stopping targets of Mu2e and COMET.}

\section{Rates and luminosity limits}

The redshifted $\chi_1$ energy stays below $m_1$, so $\chi_1$ remains bound to the star, and the volume-averaged rate per muon lies an order of magnitude below the central one, because conversion runs mainly in the inner core. The luminosity limit with which we compare these rates is $L_\infty^{\rm lim}=4\pi R_\infty^2\sigma T_\infty^4=4\pi R^2\sigma T_\infty^4/(1-2M/R)$~\cite{Guillot2019,Rodriguez2025}, with the recalibrated redshifted bound $T_\infty<3\times10^4$~K of Ref.~\cite{Rodriguez2025} and the radius and mass of each model star, which gives $1.2$--$1.8\times10^{27}$~erg~s$^{-1}$. The temperature bound comes from blackbody fits to upper limits on the far-ultraviolet flux and takes the largest temperature allowed by the uncertainties in distance, extinction and stellar radius~\cite{Guillot2019,Rodriguez2025}. In Ref.~\cite{Guillot2019} this largest temperature occurs at the smallest radius, $10$~km, so combining the bound with the larger radii of our model stars overestimates $L_\infty^{\rm lim}$ and keeps the stellar bound conservative. For the magnetic fields of ordinary pulsars the ultraviolet tail of hydrogen atmospheres approaches the Rayleigh-Jeans tail of a blackbody, whereas some models of condensed-surface emission lie below it and would allow a warmer surface~\cite{Rodriguez2025}.

Since $L_\infty\propto G^2$, the bound is $G\le G_{\rm b}(L_\infty^{\rm lim}/L_{\rm b})^{1/2}$, where the subscript b denotes benchmark~I, and benchmarks I and II of Ref.~\cite{Fox2024} share the coupling $G=(300~{\rm TeV})^{-2}$ and the masses. We map the stellar rate to $R$ with our aluminium rates at this coupling (Sec.~\ref{sec:al}), $R=4.6\times10^{-16}$ for benchmark~I, three-body, and $2.5\times10^{-18}$ for benchmark~II, on shell, instead of the estimate $3.3\times10^{-15}$ of Ref.~\cite{Fox2024} for both. The single-event sensitivities in Fig.~1 of the Letter are those quoted for COMET Phases~I and II~\cite{COMET}, for Mu2e and for Mu2e-II~\cite{Mu2eII}, and the design reports give values within thirty percent of these~\cite{COMETTDR,Mu2e}. Neutrino cooling is negligible at the temperature bound, as expected in the photon cooling stage of old neutron stars~\cite{YakovlevPethick2004}: the iron-envelope relation of Ref.~\cite{YakovlevPethick2004}, extrapolated to this temperature, gives a core temperature of a few times $10^5$~K, a light-element envelope gives a lower one, and at this temperature even direct Urca emission throughout the core~\cite{Yakovlev2001} stays more than two orders of magnitude below the luminosity limit. Pauli blocking of the dark states is negligible in the heating regime, where only a tiny fraction of the star converts.

For the scan in the $\chi_2$ mass we kept $m_{\chi_1}=m_{\chi_2}-130$~MeV and $m_{A'}=20$~MeV, a conservative choice, since a larger splitting deposits more energy in the $A'$. The lower end of the laboratory window, $m_{\chi_2}\simeq943$~MeV, is the kinematic condition $m_{\chi_1}+m_{A'}>m_p-m_\mu$ that forbids proton decay with an on-shell muon and $A'$ (Table~I of Ref.~\cite{Fox2024}). Above this edge the proton can still decay through a virtual muon, with a lifetime that Ref.~\cite{Fox2024} finds to rise steeply with $m_{\chi_1}$ and to lie above current limits only near the benchmark masses, so the window is set by kinematics and proton decay may close part of its lower end. The bound for each star ends exactly at $m_{\chi_2}=\mu_p+\mu_\mu=\mu_n$ at its centre, which rises from about $1.1$~GeV at $1.2\,\msun$ to $1.6$--$2.0$~GeV at the maximum mass. For PSR J0437$-$4715 we used $M=1.418\,\msun$~\cite{Reardon2024} and $T_\infty=2.66\times10^5$~K, twice the quoted uncertainty above the central value of the hydrogen-atmosphere fit of Ref.~\cite{Rodriguez2025} to its ultraviolet and soft X-ray spectra, because that reference does not state the confidence level of the interval and earlier fits to the far-ultraviolet emission gave a wide range of temperatures~\cite{Durant2012}. This weakens the bound on $\Lambda$ by about an order of magnitude relative to a $1.4\,\msun$ star with the limit of PSR J2144$-$3933.

\section{Off-shell channel}

With the vector vertex $g_D\bar\chi_2\gamma^\mu\chi_1A'_\mu$ of benchmark~I, the reaction $\mu^-p\to\chi_0\chi_1A'$ proceeds through a virtual $\chi_2$ of four-momentum $q=p_1+k$, where $p_1$ and $k$ are the momenta of $\chi_1$ and $A'$. Summed over all spins and $A'$ polarizations,
\begin{equation}
\sum|\mathcal M|^2=\frac{G^2}{2}\,g_D^2\,\frac{4\,(p_0\cdot p_\mu)\,T}{(q^2-m_2^2)^2+m_2^2\Gamma_2^2},\qquad
T=4\,p_1\cdot V-12\,m_1S+\frac{8\,(k\cdot V)(k\cdot p_1)}{m_{A'}^2},
\end{equation}
with $V=2(q\cdot p)\,q+(m_2^2-q^2)\,p+2Mm_2\,q$ and $S=2m_2\,q\cdot p+M(q^2+m_2^2)$, where $p$ and $M$ are the kinetic momentum and the Dirac mass of the proton. The last term of $T$ comes from the longitudinal $A'$ and dominates because the off-diagonal current is not conserved. We checked $T$ against explicit Dirac traces at random off-shell kinematics, and the same vertex gives the $\chi_2$ width of Eq.~(19) of Ref.~\cite{Fox2024} exactly, $\Gamma_2=8.7$~MeV for benchmark~I. We sampled the three-body phase space as $d\Phi_3=d\Phi_2(\sqrt s;m_0,m_Z)\,(dm_Z^2/2\pi)\,d\Phi_2(m_Z;m_1,m_{A'})$ with a Breit-Wigner mapping of $m_Z^2$, and for a narrow $\chi_2$ the three-body rate reproduces the two-body rate times the branching ratio, in the core and in aluminium.

For the stellar rate we used the Monte Carlo integration of Sec.~\ref{sec:rate} with this final state and the heat $E_{A'}$ plus the hole energy. The transition dipole of benchmark~II gives a far narrower $\chi_2$, and its off-shell contribution is negligible. Above the endpoint of the stellar on-shell channel both rates are off shell and scale as $\alpha_DG^2$, so there the ratio of the two bounds on $\Lambda$ does not depend on $\alpha_D$. The $A'$ energy is at most $W-m_1-m_0$, which falls below the energy of the conversion electron, about $105$~MeV, for $m_2\gtrsim1060$~MeV, so the Mu2e reach in Fig.~2 of the Letter, which assumes full acceptance, is optimistic there. It is optimistic at the benchmark mass as well, because the Mu2e single-event sensitivity contains the small acceptance for conversion electrons, which must leave the target with large transverse momentum~\cite{Mu2e}, while a lepton from $A'\to e^+e^-$ exceeds the conversion energy, above the decay-in-orbit background~\cite{Fox2024}, only when the $A'$ decays close to its direction of flight.

\section{Exotic capture in aluminium}\label{sec:al}

We computed the exotic capture rate in the impulse approximation, with the muon in the $1s$ state of the potential of a Fermi charge distribution with the measured radius of $^{27}$Al and with the reduced mass. The protons occupy the $1s_{1/2}$, $1p_{3/2}$, $1p_{1/2}$ and $1d_{5/2}$ orbitals of a Woods-Saxon potential with the radius, diffuseness and spin-orbit strength of Bohr and Mottelson~\cite{BohrMottelson}, close to the set listed as Wahlborn in Ref.~\cite{Schwierz2007}, and with the Coulomb potential of a uniformly charged sphere, and we fitted the depth to the separation energy $S_p=8.27$~MeV of the $1d_{5/2}$ proton. Capture of a proton from orbital $\alpha$ leaves $^{26}$Mg with a hole of energy $S_\alpha-S_p$ that recoils with momentum $-\bm P$, so
\begin{equation}
\Gamma=\sum_\alpha N_\alpha\int\frac{d^3P}{(2\pi)^3}\,\overline{|F_\alpha(\bm P)|^2}\,\frac{1}{4E_\mu E_p}\int d\Phi\,\frac14\sum|\mathcal M|^2 ,\qquad
F_\alpha(\bm P)=\int d^3r\,e^{-i\bm P\cdot\bm r}\psi_\mu(\bm r)\phi_\alpha(\bm r),
\end{equation}
where $N_\alpha$ is the occupation, the bar denotes the average over the magnetic substates, and the dark final state has total momentum $\bm P$ and energy $W_\alpha=M(\mu^{27}{\rm Al})-M_\alpha-P^2/2M_\alpha$, with $M_\alpha$ the mass of $^{26}$Mg with the hole. The proton four-momentum is then $(W_\alpha-E_\mu,\bm P)$, the overlaps satisfy Parseval's relation, and we divide the rate by the measured total capture rate in aluminium~\cite{Suzuki1987}. As a test of the normalization we computed ordinary capture, $\mu^-p\to\nu_\mu n$, with the same code and the $V-A$ matrix element without induced couplings. For muonic hydrogen it gives a statistical rate six percent above the lower bound $\Lambda_S/4$ that the measured singlet rate $\Lambda_S$~\cite{MuCap2013} sets, and for aluminium without Pauli blocking of the neutron it gives three times the measured rate, as expected for the impulse approximation. An independent implementation with a harmonic-oscillator orbital gives on-shell rates close to ours.

With this normalization and at $G=(300~{\rm TeV})^{-2}$, the on-shell rate at the benchmark mass is $R=2.5\times10^{-18}$, to which only the $1d_{5/2}$ protons contribute, and the estimate of Ref.~\cite{Fox2024} exceeds our rate by a factor that grows from four at $m_2=945$~MeV to more than a thousand at $1030$~MeV. Near threshold the rate falls as the fourth to fifth power of the energy release $W-m_2$ rather than the square, because the Dalitz-area ratio of Ref.~\cite{Fox2024} overestimates the exact phase space and the matrix element $\propto p_0\cdot p_\mu=m_\mu E_0$ vanishes with the energy of the massless $\chi_0$. With the physical width of benchmark~I, the three-body rate through a virtual $\chi_2$ reaches all proton shells and gives $R=4.6\times10^{-16}$ at the benchmark mass, nearly two hundred times the on-shell rate, of which the $1d_{5/2}$ shell supplies half. The three-body rate falls steeply above the on-shell threshold, and the channel closes at $1145$~MeV, where $m_{\chi_1}+m_{A'}$ reaches the energy release.

These rates depend on how the proton-hole strength is distributed over the states of $^{26}$Mg, which is the main nuclear uncertainty. Moving the $1d_{5/2}$ strength $2$~MeV up lowers the on-shell rate at the benchmark mass by an order of magnitude but changes it further below threshold by only a quarter, admixture of $2s_{1/2}$ strength in low-lying states raises the rate near threshold, and quenching of spectroscopic factors lowers it. We therefore assign a factor of three to the on-shell rate at the benchmark mass and a factor of $1.5$ to the three-body rate and to the on-shell rate further below threshold, and even the upper end leaves benchmark~II below the Mu2e single-event sensitivity.

\section{Non-relativistic EoS check}

We built cold $npe\mu$ matter in beta equilibrium from the Skyrme force SLy4~\cite{Chabanat1998}, with the parameters of Ref.~\cite{Sun2024}, joined to the same crust. It gives $M_{\max}=2.05\,\msun$ and $R_{1.4}=11.7$~km, the standard values for SLy, whereas the radius tabulated in Ref.~\cite{Sun2024} refers to a stellar model with pure-neutron-matter cores. Its central neutron chemical potential lies above the RMF range and above the threshold for capture in aluminium. For the rate we used the Skyrme Landau mass of the proton and fixed its vector potential by $\mu_p$. With our aluminium rate for benchmark~II, a $1.2\,\msun$ SLy4 star gives the heating bound $R<1.7\times10^{-36}$ at the benchmark masses, an order of magnitude stronger than the most conservative RMF model.

\section{Annihilation and burn-down}

For a Majorana $\chi_1$ with $t$- and $u$-channel $\chi_2$ exchange we estimated
\begin{equation}
\sigma v\simeq\frac{\pi\alpha_D^2}{m_1^2}\frac{(1-r)^{3/2}}{(1-r/2)^2}\left[\frac{2m_1^2}{m_1^2+m_2^2-m_{A'}^2}\right]^2,\qquad r=\frac{m_{A'}^2}{m_1^2},
\end{equation}
and $\sigma v\sim d^4m_1^2/\pi$ for the dipole benchmark\revc{, which applies when $\chi_1$ is a Majorana fermion. Reference~\cite{Fox2024} allows the transition dipole between Dirac or Majorana fermions; the Letter takes benchmark~II as the number-conserving Dirac case, and a Majorana $\chi_1$ with a dipole coupling annihilates and falls under the bounds of benchmark~I}. \rev{These estimates exceed the cross section needed for annihilation to keep pace with production by many orders of magnitude, so the steady-state $\chi_1$ density is far below the density that would Pauli block conversion, and factor-of-few errors in the estimates do not matter.} \revb{Reference~\cite{Fox2024} treats the massive dark fermions of its benchmarks as stable and retained by the star, but annihilation removes them, and the} star therefore \revb{moves quasi-statically along its nucleonic sequence to lower baryonic mass, as in the treatment of baryon loss with escaping dark products of Ref.~\cite{Berryman2024}, and} burns baryons until $\mu_{n,c}$ falls below $m_{\chi_2}$, which leaves a star lighter than any of $1.174\,\msun$, since $\mu_{n,c}$ of such a star exceeds the benchmark $\chi_2$ mass in every model. The burn-down time from $1.4$ to $1.174\,\msun$ uses the volume-integrated rates of $1.2$ and $1.4\,\msun$ stars, interpolated logarithmically in mass, and is $0.9$--$4.5$~yr at the benchmark coupling.

\section{Dark equilibrium, collapse and muon regeneration}\label{sec:collapse}

\begin{figure}[h]
\includegraphics[width=0.55\columnwidth]{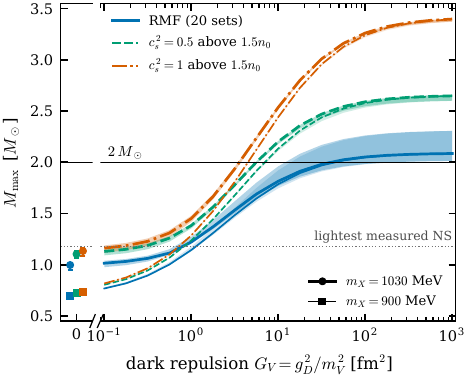}
\caption{Maximum mass of a neutron star in chemical equilibrium with a dark baryon of mass $m_X=1030$~MeV (thick lines) and $900$~MeV (thin lines), against the dark repulsion $\GV$. Lines are medians and bands span each EoS class: twenty RMF sets (solid), and the same sets continued above $1.5\,n_0$ with $c_s^2=0.5$ (dashed) or $1$ (dash-dotted). The points at the left give $\GV=0$. The dotted line marks the $1.174\,\msun$ companion of PSR J0453+1559.}
\label{fig:mmax}
\end{figure}

In Fig.~\ref{fig:mmax} we show the maximum mass in dark equilibrium against $\GV$. Without repulsion the single-species maximum mass lies below $1.174\,\msun$ for all sixty EoSs and both dark-baryon masses, and the two-species end state of the benchmark models, with $\chi_2$ and $\chi_1$ seas both filled to $\mu_B$ (the decay $\chi_2\to\chi_1A'$ with an escaping or decaying $A'$ gives $\mu_1=\mu_2$), supports only about $0.6\,\msun$.

Reference~\cite{Fox2024} also proposes that $\chi_1$ cascades into several lighter dark fermions, which would change this end state. For a cascade into three fermions $\chi_3$ of baryon number $1/3$, equilibrium sets $\mu_3=\mu_B/3$, and since the pressure and density of a free Fermi gas scale as $P=m^4f(\mu/m)$ and $n=m^3g(\mu/m)$, a cascade with $m_3$ close to $m_1/3$ gives at each $\mu_B$ a dark pressure $1/81$ of that of a single $\chi_1$ sea and absorbs $1/81$ as many baryons. The softening is then much weaker, and the star may survive without repulsion\revc{, as studies of the neutron decay into three dark fermions of baryon number $1/3$ find: with $\mu_\chi=\mu_n/3$ the equation of state stays nearly as stiff as without dark fermions, and stars of about $2\,\msun$ remain possible, even with several such fermions~\cite{Strumia2022,Darini2023}}. A lighter $\chi_3$ absorbs more baryons, and at the benchmark masses $3m_3$ lies below the baryon chemical potential at the surface, which is close to the nucleon mass, so the dark gas also extends beyond the nucleons, which our single-fluid tables do not describe. We have not computed these stars, so the cascade remains a second way for a signal at Mu2e or COMET to survive.

To find when a star with a thermalized dark sea collapses, we computed the critical converted fraction by filling the dark sea up to $\mu_B-\delta$ and lowering $\delta$ until the maximum baryonic mass falls below that of a $1.4\,\msun$ star. It is $2$--$3\%$ for $m_X=1030$~MeV and $4$--$6\%$ for $900$~MeV, and Fig.~1 of the Letter uses the larger \revb{value. Full conversion of any star with more baryons than the dark-equilibrium maximum ends in a black hole, as Ref.~\cite{Motta2018} found for dark neutron decay from the same baryon-number argument; the critical fraction shows that collapse sets in long before the dark sea fills}. With the volume-integrated rate, collapse at the benchmark coupling takes $14$--$84$ days. Since every conversion consumes a muon and a proton, we checked that the composition stays near equilibrium. In the heating regime the consumption is irrelevant, because at the bound rate a star consumes less than $10^{-8}$ of its muons over the age of PSR J2144$-$3933, so the composition stays in equilibrium without regeneration, and the fractional baryon loss lies six orders of magnitude below the limit that the orbital period of the double pulsar J0737$-$3039A/B sets~\cite{Berryman2024}. In the collapse regime weak interactions must regenerate muons and protons; the modified Urca relaxation times of Ref.~\cite{Yakovlev2001}, evaluated in the suprathermal regime where the chemical imbalance greatly exceeds the temperature, are at most about a second, far shorter than the days over which the star converts, so the muon population stays near equilibrium. Nucleon superfluid gaps, of order an MeV at most~\cite{Yakovlev2001}, are small compared with the tens of MeV of phase space in the conversion and do not affect these estimates.

\rev{\section{Thermalization and the collisionless end state}

\emph{Energy loss.} The benchmark dark sector has no tree-level elastic self-interaction, and $\chi_1$ couples to charged matter only through the off-diagonal \revb{vertex, which Ref.~\cite{Fox2024} requires in both benchmarks}. Elastic scattering on a proton or an electron therefore first appears at second order: $\chi_1$ absorbs a dark photon, becomes a virtual $\chi_2$, and returns by emitting another. For energy transfers small compared with $\Delta=m_2-m_1$ this is the potential $V_{\rm el}(r)=-V_{12}(r)^2/\Delta$, built from the transition potential
\begin{equation}
V_{12}(q)=\frac{\varepsilon e\,g\,q^2}{(q^2+m_{A'}^2)(q^2+k_D^2)} ,
\end{equation}
where the factor $q^2/(q^2+k_D^2)$ is the screening of the kinetic mixing by the plasma, with the Debye momentum $k_D$ of the degenerate electrons, muons and \revb{protons, $k_D^2=\sum_i3\omega_{p,i}^2/v_{F,i}^2$, the static limit of the longitudinal polarization~\cite{BraatenSegel1993}}. For the vector coupling $g=g_D$; for the transition dipole of benchmark~II we set $g=d_{\rm tr}|q|$ and dropped the velocity suppression of the dipole vertex, which overestimates the coupling. A $\chi_1$ with velocity $\bm v$ loses energy at the rate
\begin{equation}
\frac{dE}{dt}=\int\frac{d^3q}{(2\pi)^3}\,|\tilde V_{\rm el}(q)|^2\,\omega\,S(q,\omega),\qquad \omega=\bm q\cdot\bm v-\frac{q^2}{2m_1},
\end{equation}
with the zero-temperature Lindhard structure factor $S$ of the protons and electrons, using the Fermi energy as Landau mass. In vacuum the transform $\tilde V_{\rm el}$ agrees with the analytic form $-(4\pi C/q)\arctan(q/2m_{A'})$, $C=(\varepsilon eg/4\pi)^2/\Delta$, and $S$ satisfies the $f$-sum rule. We integrated from the production momentum down to a kinetic energy of $0.1$~MeV at the centres of $1.2$ and $1.4\,\msun$ stars. For benchmark~I at $\varepsilon=10^{-4}$ this takes $(1.2$--$2.0)\times10^7$~s; the time scales as $\varepsilon^{-4}\alpha_D^{-2}$, so $\chi_1$ cools within $1$~Gyr for $\varepsilon\gtrsim5\times10^{-7}$. For benchmark~II it takes $(1.6$--$3.7)\times10^{16}$~s even with the overestimated coupling, much longer than conversion takes at the rates of interest (see below). With $g=d_{\rm tr}|q|$ the potential behaves as $r^{-4}$ at short distance and its transform depends on the momentum cutoff, and the physical cutoff $q\approx\sqrt{2m_1\Delta}$, lower than ours, gives a still longer time.

\emph{Collisionless gas.} \revb{Earlier studies of dark neutron decay assumed a degenerate dark sea in equilibrium with the nucleons~\cite{McKeen2018,Baym2018,Motta2018}; Ref.~\cite{Husain2022} noted that the dark particles are born far above the bottom of the sea and cool slowly, and treated them at a finite temperature. Without} energy loss each $\chi_1$ keeps its Killing energy $E_\infty=e^\Phi E$ in a static star. Production feeds the orbits from the bottom of the potential, $E_\infty=m_1e^{\Phi_c}$, up to energies well below $m_1$. Production into a phase-space cell of occupation $F$ is reduced by $1-F$, so the occupation is $F=1-\exp(-A s)$, where $s$ is the production rate per unit phase-space volume, normalized to its maximum, and $A$ grows linearly with time. Since the star contracts slowly as conversion proceeds, we froze the occupation as a function of the phase-space volume below the orbit energy, $\Gamma(E_\infty)=\int dV\,p^3/3\pi^2$, the adiabatic invariant of an isotropic distribution, so that the dark number $\int F\,d\Gamma$ is conserved; the exact invariants are the orbital actions, so this is an approximation. We solved the TOV equations for the nucleons and this dark gas, whose density, energy density and pressure at each radius follow from $F$, iterating between the lapse profile and $\Gamma(E_\infty)$ and shooting on the central lapse to match the exterior metric. The energy grid covers all bound orbits, $E_\infty<m_1$, because a compact configuration has a deeper potential but less volume, and it holds a given $\Gamma$ only at higher $E_\infty$. We accepted a configuration only if the iteration converged, and if it holds the conserved dark number; configurations that cannot bind the dark gas are thereby excluded. Without dark gas the solver reproduces the $1.4\,\msun$ star, and for small $A$ the largest baryon number approaches that of the nucleonic maximum-mass star. The star collapses when the largest baryon number over central pressures falls below that of the original star. In this isotropic approximation FSU2R and GM3 both collapse when about a fifth of the baryons have converted.

\emph{Orbital actions.} The exact adiabatic invariants of an orbit in a slowly contracting spherical star are its angular momentum $L$ and radial action $J_r=\pi^{-1}\int p_r\,dr$, with the covariant radial momentum $p_r=e^{\lambda}(E_\infty^2e^{-2\Phi}-m_1^2-L^2/r^2)^{1/2}$. We therefore repeated the calculation with the production rate binned in $(E_\infty,L)$, mapped to $(J_r,L)$ in the original star, and the occupation $F=1-\exp(-As)$ frozen in $(J_r,L)$. The phase-space measure is $d^3x\,d^3p=(2\pi)^3\,2L\,dJ_r\,dL$ per unit azimuth range; we checked it symbolically and by counting particles both ways in a test metric. The gas is then anisotropic, so its radial pressure enters $d\Phi/dr$, and its local density, energy density and radial pressure follow from $d^3p=4\pi p_\perp dp_\perp dp_r$ with $p_\perp=L/r$; we iterated metric, actions, occupation and the TOV equations to convergence, and the dark number is conserved to the accuracy of the binning. The production in $(J_r,L)$ can fill orbits holding $35\%$ of the baryons for FSU2R. Interpolating the largest baryon number between the converted fractions we computed, the star collapses when $21.3\%$ of its baryons have converted for FSU2R and $21.0\%$ for GM3, as in the isotropic approximation. Rejected configurations near the maximum miss the convergence cut only narrowly; iterating them to convergence leaves every unstable case unstable. A cold dark sea in the same solver collapses after a few percent, as in Sec.~\ref{sec:collapse}.

\emph{Heating while the orbits fill.} The conversion rate falls by the factor $B(A)=\int s\,e^{-As}d\Gamma/\int s\,d\Gamma$, which decreases roughly as $1/A$ from $A=10$ to $10^6$ because the production rate per orbit spans many decades. At the benchmark coupling collapse takes about two centuries, and with the on-shell aluminium rate of benchmark~II it occurs within $1$~Gyr for $R\gtrsim6\times10^{-25}$. At slower rates the star is still converting at the age of PSR J2144$-$3933, and its luminosity exceeds the limit by at least an order of magnitude at every rate between the conservative heating bound and the collapse threshold. These numbers use FSU2R at $1.4\,\msun$, the critical fraction is checked for FSU2R and GM3, and we have not repeated the collisionless calculation for the other eighteen models; we extrapolated $B$ beyond $A=10^6$ as $A^{-2}$, the behaviour near a kinematic edge, which is conservative.}

\section{The repulsion scaling law}

\begin{figure}[h]
\includegraphics[width=0.55\columnwidth]{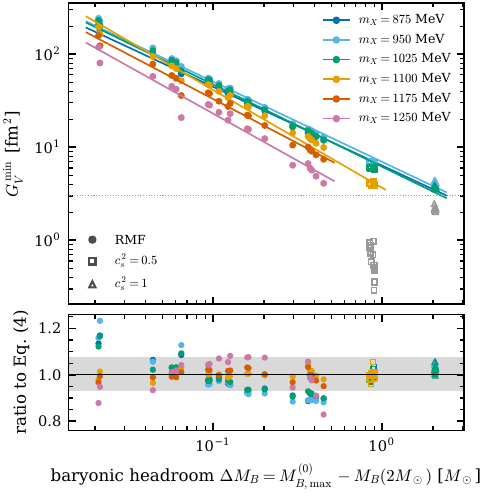}
\caption{Upper panel: minimum dark repulsion $\GVmin$ for which the converted EoS supports the baryonic mass of a $2\,\msun$ star, against the baryonic headroom $\DMB$ of the nucleonic EoS, for sixty EoSs and six dark-baryon masses. Filled circles are RMF sets; open squares and triangles are the same sets continued with $c_s^2=0.5$ and $1$. Lines are power-law fits in $\DMB$. Grey points lie below $3$~fm$^2$ (dotted line), outside the domain of the law. Lower panel: ratio of $\GVmin$ to Eq.~(4) of the Letter; the grey band marks $\pm$ the largest root-mean-square scatter over all sixteen masses.}
\label{fig:law}
\end{figure}

In Fig.~\ref{fig:law} we show $\GVmin$ against the baryonic headroom for six of the sixteen dark-baryon masses. The EoSs fall on one line for each mass, with an exponent close to the value $a=1$ suggested by the argument in the Letter. All sixty EoSs enter the fit up to $m_X=1050$~MeV; at heavier masses the stiff extensions leave the domain of the law. The chemical-potential factor removes the remaining offset between the EoS classes, and the scatter about Eq.~(4) of the Letter is two to eight percent at every mass. \rev{The exponent $b$ rises with $m_X$ (Table~\ref{tab:law}), so} the simple argument identifies the variables but not their full dependence. The law fails only where the requirement is a few fm$^2$ or less, for a heavy dark baryon and a stiff EoS, because the converted star then almost reaches $2\,\msun$ without repulsion; these EoSs appear as grey \revb{points. Earlier studies of a repulsive dark baryon near the neutron mass gave the requirement for one to three EoSs at a time, from the condition that the converted maximum gravitational mass reach $2\,\msun$, and in our notation they found values between about ten and sixty fm$^2$~\cite{Cline2018,Grinstein2019,Husain2022}, with the softer EoS needing the larger repulsion, as Eq.~(4) of the Letter implies. Reference~\cite{Husain2022} also followed a star of fixed baryon number as $\GV$ varies, for one EoS, and Ref.~\cite{Grinstein2019} mainly considered a dark-baryon--neutron repulsion, which Eq.~(4) does not describe; none of these studies relates the requirement to the headroom of the nucleonic EoS. A recent scan of neutron conversion to a repulsive dark fermion over twenty-one RMF EoSs~\cite{RafieiKarkevandi2026} finds that weak repulsion is excluded for every EoS while the rest of the excluded region depends on the EoS, the dependence that Eq.~(4) of the Letter traces to the headroom. The gravitational headroom plays the same role for dark matter carried along with the baryons, whose limit is the headroom divided by a nearly universal rate~\cite{Kalita2026}; here the relevant quantity is the baryonic headroom, because conversion conserves baryon number}.

\section{Coefficients of the repulsion law}

Table~\ref{tab:law} lists the coefficients of Eq.~(4) of the Letter, fitted to the $N$ EoSs with $\GVmin\ge3$~fm$^2$. The jump near $1.1$~GeV marks where the extended EoSs leave the domain of the law. Without this cut the scatter grows sharply above $1.1$~GeV, while within the domain no EoS deviates from the law by more than a quarter. The extended EoSs share the RMF physics below $1.5\,n_0$ and cluster at two values of the headroom, so they test the law mainly as two lever arms, and the exponent $b$ rises across the mass grid. \rev{A law of the same form with similar coefficients holds for the gravitational-mass criterion, and for a bosonic dark baryon (Sec.~\ref{sec:lawtest}).} A fine grid in $m_X$ and an independent coarser run agree to numerical precision where they overlap.

\begin{table}[h]
\caption{Coefficients of Eq.~(4) of the Letter, number of EoSs in the fit, and root-mean-square scatter in $\log_{10}\GVmin$, for a spin-$\tfrac12$ dark baryon and the baryonic-mass criterion.}
\label{tab:law}
\begin{ruledtabular}
\begin{tabular}{cccccc}
$m_X$ (MeV) & $A$ (fm$^2$) & $a$ & $b$ & $N$ & rms (dex)\\
\hline
875 & 1.36 & 0.89 & 0.65 & 60 & 0.022\\
900 & 1.68 & 0.90 & 0.59 & 60 & 0.025\\
925 & 1.93 & 0.90 & 0.55 & 60 & 0.026\\
950 & 2.06 & 0.90 & 0.53 & 60 & 0.026\\
975 & 2.05 & 0.91 & 0.53 & 60 & 0.025\\
1000 & 1.95 & 0.92 & 0.54 & 60 & 0.024\\
1025 & 1.79 & 0.93 & 0.56 & 60 & 0.020\\
1050 & 1.60 & 0.95 & 0.58 & 60 & 0.016\\
1075 & 1.52 & 0.98 & 0.55 & 41 & 0.010\\
1100 & 1.05 & 1.02 & 0.67 & 40 & 0.008\\
1125 & 0.53 & 1.08 & 0.93 & 36 & 0.015\\
1150 & 0.85 & 1.04 & 0.71 & 20 & 0.008\\
1175 & 0.59 & 1.07 & 0.84 & 20 & 0.010\\
1200 & 0.38 & 1.10 & 0.98 & 20 & 0.014\\
1225 & 0.28 & 1.14 & 1.08 & 20 & 0.021\\
1250 & 0.17 & 1.20 & 1.22 & 20 & 0.032\\
\end{tabular}

\end{ruledtabular}
\end{table}

\rev{\section{Out-of-sample test of the law and bosonic dark baryons}\label{sec:lawtest}

We tested Eq.~(4) of the Letter, with the coefficients of Table~\ref{tab:law} and no refit, on three families of EoSs absent from the fit. The first is twenty Skyrme forces of Ref.~\cite{Sun2024} with a single density-dependent term (SLy4--SLy7, SLy9, SLy230a, SkI2, SkI3, SkI5, SK255, SK272, SkMP, SkO, Ska35s25, SkT4a, KDE0v1, NRAPR, Rs, Gs and SV-min), each truncated where it becomes acausal; their saturation properties agree with Ref.~\cite{Sun2024}, and sixteen reach $2.01\,\msun$. The second family consists of sixteen hybrid EoSs, built from FSU2R or NLM6 up to a transition density of $1.5$ or $2\,n_0$, a Maxwell transition with an energy-density jump of $0.2$ or $0.4$ times its value at the transition, and quark matter with $c_s^2=2/3$ or $1$ above \revb{it~\cite{AlfordHan2013}; both jumps lie below the threshold $\Delta\varepsilon_{\rm crit}/\varepsilon_t=1/2+3p_t/(2\varepsilon_t)$ beyond which the hybrid branch connected to the nucleonic stars is unstable~\cite{AlfordHan2013}}. For each EoS we found $\GVmin$ by bisection at $m_X=900$, $1000$, $1100$ and $1200$~MeV and kept the cases with $\GVmin\ge3$~fm$^2$, the domain of the law. In Fig.~\ref{fig:lawtest} we show the ratio of $\GVmin$ to the law. The Skyrme forces and the hybrid EoSs both follow the law with a root-mean-square scatter of eight percent, and the largest deviation, sixteen percent, occurs for a Skyrme force. The RMF set FSU2R of the fit, recomputed with the same bisection, deviates by five percent, so the method itself contributes little. The law therefore carries over to non-relativistic interactions and to strong first-order transitions, with a scatter comparable to the largest scatter of the fit.

\emph{Hyperons.} We added the baryon octet to the RMF sets of types without density-dependent couplings or a $\delta$ meson. The hyperons couple to $\sigma$, $\omega$, $\rho$ and the hidden-strangeness $\phi$ \revb{meson, as in the $\sigma\omega\rho\phi$ model of Ref.~\cite{Weissenborn2012}. The scalar couplings follow from the potentials $U_\Lambda=-30$, $U_\Sigma=+30$ and $U_\Xi=-28$~MeV in symmetric matter at saturation, the values adopted in Ref.~\cite{Weissenborn2012}}, and the $\rho$ couplings follow from isospin. With the SU(6) vector couplings no set reaches $2\,\msun$, and we therefore used the SU(3) vector couplings of Ref.~\cite{Weissenborn2012}, at $z=g_8/g_1=0$, $0.1$ and $0.2$, with the nucleon-$\phi$ coupling of that scheme set to zero so that each nucleonic EoS is unchanged; Ref.~\cite{Weissenborn2012} instead keeps this coupling and refits $g_{N\omega}$, which requires refitting nucleonic models with vector self-interactions such as ours. With the hyperons switched off, the code reproduces the nucleonic EoS to machine precision, and at $z=1/\sqrt6$ it reproduces the SU(6) couplings. Seventeen EoSs reach $2.01\,\msun$, with maximum masses up to $2.23\,\msun$, baryonic headrooms of $0.02$--$0.33\,\msun$, and hyperon fractions of $15$--$43\%$ in the centre of the maximum-mass star. For a fermionic dark baryon at $900$--$1200$~MeV they follow Eq.~(4) of the Letter, without refit, with a root-mean-square scatter of six percent and at most fourteen percent.

\emph{Bosonic dark baryons.} A bosonic dark baryon forms a condensate with $n_X=(\mu_B-m_X)/\GV$ and $P_X=\GV n_X^2/2$. For $m_X\gtrsim1$~GeV the condensate forms only in the core, and the baryonic-mass criterion behaves as for fermions. Fitting Eq.~(4) of the Letter to the sixty nucleonic EoSs gives $A=2.37$~fm$^2$, $a=0.86$ and $b=0.54$ at $m_X=1030$~MeV, and $A=0.57$~fm$^2$, $a=1.03$ and $b=0.90$ at $1200$~MeV, with root-mean-square scatters of seven and two percent. For the same EoS this is up to a third more repulsion than for a fermion, because the condensate has no Fermi pressure. The seventeen hyperonic EoSs follow this bosonic law with a root-mean-square scatter of eight percent and at most seventeen percent. At $m_X=938$~MeV the condensate forms at crust densities, and for weak repulsion the dark-matter-dominated configurations have no meaningful baryonic-mass maximum, so we exclude this case.

\begin{figure}[h]
\includegraphics[width=0.55\columnwidth]{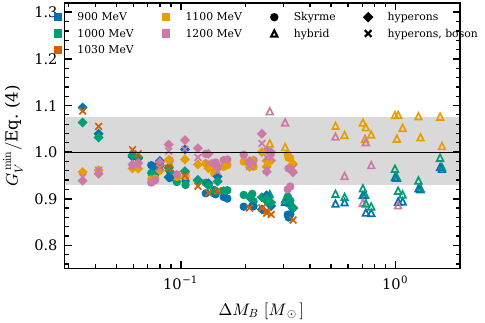}
\caption{Ratio of the minimum dark repulsion $\GVmin$ to the scaling law, Eq.~(4) of the Letter with the coefficients of Table~\ref{tab:law}, for EoSs not used in the fit: sixteen Skyrme forces (filled circles), sixteen hybrid EoSs with a first-order transition to quark matter (open triangles) and seventeen hyperonic EoSs (diamonds), at four dark-baryon masses. Crosses: bosonic dark baryon on the hyperonic EoSs, against the bosonic law at $1030$ and $1200$~MeV. The grey band marks $\pm$ the largest scatter of the fit.}
\label{fig:lawtest}
\end{figure}}


\begin{thebibliography}{0}%
\makeatletter
\providecommand \@ifxundefined [1]{%
 \@ifx{#1\undefined}
}%
\providecommand \@ifnum [1]{%
 \ifnum #1\expandafter \@firstoftwo
 \else \expandafter \@secondoftwo
 \fi
}%
\providecommand \@ifx [1]{%
 \ifx #1\expandafter \@firstoftwo
 \else \expandafter \@secondoftwo
 \fi
}%
\providecommand \natexlab [1]{#1}%
\providecommand \enquote  [1]{``#1''}%
\providecommand \bibnamefont  [1]{#1}%
\providecommand \bibfnamefont [1]{#1}%
\providecommand \citenamefont [1]{#1}%
\providecommand \href@noop [0]{\@secondoftwo}%
\providecommand \href [0]{\begingroup \@sanitize@url \@href}%
\providecommand \@href[1]{\@@startlink{#1}\@@href}%
\providecommand \@@href[1]{\endgroup#1\@@endlink}%
\providecommand \@sanitize@url [0]{\catcode `\\12\catcode `\$12\catcode
  `\&12\catcode `\#12\catcode `\^12\catcode `\_12\catcode `\%12\relax}%
\providecommand \@@startlink[1]{}%
\providecommand \@@endlink[0]{}%
\providecommand \url  [0]{\begingroup\@sanitize@url \@url }%
\providecommand \@url [1]{\endgroup\@href {#1}{\urlprefix }}%
\providecommand \urlprefix  [0]{URL }%
\providecommand \Eprint [0]{\href }%
\providecommand \doibase [0]{https://doi.org/}%
\providecommand \selectlanguage [0]{\@gobble}%
\providecommand \bibinfo  [0]{\@secondoftwo}%
\providecommand \bibfield  [0]{\@secondoftwo}%
\providecommand \translation [1]{[#1]}%
\providecommand \BibitemOpen [0]{}%
\providecommand \bibitemStop [0]{}%
\providecommand \bibitemNoStop [0]{.\EOS\space}%
\providecommand \EOS [0]{\spacefactor3000\relax}%
\providecommand \BibitemShut  [1]{\csname bibitem#1\endcsname}%
\let\auto@bib@innerbib\@empty
\end{thebibliography}%


\begin{thebibliography}{99}
\bibitem{Sakharov1967} A. D. Sakharov, Violation of CP invariance, C asymmetry, and baryon asymmetry of the universe, JETP Lett. \textbf{5}, 24 (1967).
\bibitem{SuperK2020} A. Takenaka \textit{et al.} (Super-Kamiokande Collaboration), Search for proton decay via $p\to e^+\pi^0$ and $p\to\mu^+\pi^0$ with an enlarged fiducial volume in Super-Kamiokande I-IV, Phys. Rev. D \textbf{102}, 112011 (2020).
\bibitem{KamLAND2006} T. Araki \textit{et al.} (KamLAND Collaboration), Search for the invisible decay of neutrons with KamLAND, Phys. Rev. Lett. \textbf{96}, 101802 (2006).
\bibitem{Fox2024} P. J. Fox, M. Hostert, T. Menzo, M. Pospelov, and J. Zupan, Muon-induced baryon number violation, Phys. Rev. D \textbf{110}, 075015 (2024).
\bibitem{Mu2e} L. Bartoszek \textit{et al.} (Mu2e Collaboration), Mu2e technical design report, \href{https://arxiv.org/abs/1501.05241}{arXiv:1501.05241}.
\bibitem{Mu2eII} F. Abusalma \textit{et al.} (Mu2e Collaboration), Expression of interest for evolution of the Mu2e experiment, \href{https://arxiv.org/abs/1802.02599}{arXiv:1802.02599}.
\bibitem{COMETTDR} R. Abramishvili \textit{et al.} (COMET Collaboration), COMET Phase-I technical design report, Prog. Theor. Exp. Phys. \textbf{2020}, 033C01 (2020).
\bibitem{COMET} M. Moritsu (COMET Collaboration), Search for muon-to-electron conversion with the COMET experiment, Universe \textbf{8}, 196 (2022).
\bibitem{Calibbi2018} L. Calibbi and G. Signorelli, Charged lepton flavour violation: An experimental and theoretical introduction, Riv. Nuovo Cim. \textbf{41}, 71 (2018).
\bibitem{ZhangLi2020} N.-B. Zhang and B.-A. Li, Constraints on the muon fraction and density profile in neutron stars, Astrophys. J. \textbf{893}, 61 (2020).
\bibitem{SM} See Supplemental Material for the stellar structure and its validation, the non-relativistic EoS check, the conversion rate and heat, the in-medium dark photon, annihilation and burn-down, the critical converted fraction and muon regeneration, the maximum mass against the dark repulsion, and the scaling law, which includes Ref.~\cite{Yakovlev2001}.
\bibitem{Yakovlev2001} D. G. Yakovlev, A. D. Kaminker, O. Y. Gnedin, and P. Haensel, Neutrino emission from neutron stars, Phys. Rep. \textbf{354}, 1 (2001).
\bibitem{Bollig2020} R. Bollig, W. DeRocco, P. W. Graham, and H.-T. Janka, Muons in supernovae: implications for the axion-muon coupling, Phys. Rev. Lett. \textbf{125}, 051104 (2020).
\bibitem{Fiorillo2026} D. F. G. Fiorillo, A. Lella, G. G. Raffelt, N. Selimovic, and E. Vitagliano, Neutron star bounds on muonic fifth forces from picometer to meter scales, \href{https://arxiv.org/abs/2605.24094}{arXiv:2605.24094}.
\bibitem{McKeen2021} D. McKeen, M. Pospelov, and N. Raj, Neutron star internal heating constraints on mirror matter, Phys. Rev. Lett. \textbf{127}, 061805 (2021).
\bibitem{Guillot2019} S. Guillot, G. G. Pavlov, C. Reyes, A. Reisenegger, L. Rodriguez, B. Rangelov, and O. Kargaltsev, Hubble Space Telescope nondetection of PSR J2144$-$3933: the coldest known neutron star, Astrophys. J. \textbf{874}, 175 (2019).
\bibitem{McKeen2018} D. McKeen, A. E. Nelson, S. Reddy, and D. Zhou, Neutron stars exclude light dark baryons, Phys. Rev. Lett. \textbf{121}, 061802 (2018).
\bibitem{Baym2018} G. Baym, D. H. Beck, P. Geltenbort, and J. Shelton, Testing dark decays of baryons in neutron stars, Phys. Rev. Lett. \textbf{121}, 061801 (2018).
\bibitem{Motta2018} T. F. Motta, P. A. M. Guichon, and A. W. Thomas, Implications of neutron star properties for the existence of light dark matter, J. Phys. G \textbf{45}, 05LT01 (2018).
\bibitem{FornalGrinstein2018} B. Fornal and B. Grinstein, Dark matter interpretation of the neutron decay anomaly, Phys. Rev. Lett. \textbf{120}, 191801 (2018).
\bibitem{Sun2024} B. Sun, S. Bhattiprolu, and J. M. Lattimer, Compiled properties of nucleonic matter and nuclear and neutron star models from non-relativistic and relativistic interactions, Phys. Rev. C \textbf{109}, 055801 (2024).
\bibitem{Fonseca2021} E. Fonseca \textit{et al.}, Refined mass and geometric measurements of the high-mass PSR J0740+6620, Astrophys. J. Lett. \textbf{915}, L12 (2021).
\bibitem{Abbott2019} B. P. Abbott \textit{et al.} (LIGO Scientific and Virgo Collaborations), Properties of the binary neutron star merger GW170817, Phys. Rev. X \textbf{9}, 011001 (2019).
\bibitem{GM1991} N. K. Glendenning and S. A. Moszkowski, Reconciliation of neutron-star masses and binding of the $\Lambda$ in hypernuclei, Phys. Rev. Lett. \textbf{67}, 2414 (1991).
\bibitem{Chabanat1998} E. Chabanat, P. Bonche, P. Haensel, J. Meyer, and R. Schaeffer, A Skyrme parametrization from subnuclear to neutron star densities. Part II. Nuclei far from stabilities, Nucl. Phys. A \textbf{635}, 231 (1998).
\bibitem{Alford2021} M. G. Alford, A. Haber, S. P. Harris, and Z. Zhang, Beta equilibrium under neutron star merger conditions, Universe \textbf{7}, 399 (2021).
\bibitem{Alford2024} M. G. Alford, A. Haber, and Z. Zhang, Isospin equilibration in neutron star mergers, Phys. Rev. C \textbf{109}, 055803 (2024).
\bibitem{An2013} H. An, M. Pospelov, and J. Pradler, New stellar constraints on dark photons, Phys. Lett. B \textbf{725}, 190 (2013).
\bibitem{BraatenSegel1993} E. Braaten and D. Segel, Neutrino energy loss from the plasma process at all temperatures and densities, Phys. Rev. D \textbf{48}, 1478 (1993).
\bibitem{Cline2018} J. M. Cline and J. M. Cornell, Dark decay of the neutron, J. High Energy Phys. \textbf{07} (2018) 081.
\bibitem{Strumia2022} A. Strumia, Dark Matter interpretation of the neutron decay anomaly, J. High Energy Phys. \textbf{02} (2022) 067.
\bibitem{Darini2023} L. Darini, Dark matter implications of the neutron anomaly, Eur. Phys. J. C \textbf{83}, 364 (2023).
\bibitem{Rodriguez2025} L. E. Rodr\'iguez, A. Reisenegger, D. Gonz\'alez-Caniulef, C. Petrovich, G. Pavlov, S. Guillot, O. Kargaltsev, and B. Rangelov, Contrasting neutron star heating mechanisms with Hubble Space Telescope observations, Astron. Astrophys. \textbf{708}, A377 (2026).
\bibitem{YakovlevPethick2004} D. G. Yakovlev and C. J. Pethick, Neutron star cooling, Annu. Rev. Astron. Astrophys. \textbf{42}, 169 (2004).
\bibitem{Reisenegger1995} A. Reisenegger, Deviations from chemical equilibrium due to spin-down as an internal heat source in neutron stars, Astrophys. J. \textbf{442}, 749 (1995).
\bibitem{Baryakhtar2017} M. Baryakhtar, J. Bramante, S. W. Li, T. Linden, and N. Raj, Dark kinetic heating of neutron stars and an infrared window on WIMPs, SIMPs, and pure Higgsinos, Phys. Rev. Lett. \textbf{119}, 131801 (2017).
\bibitem{Ema2025} Y. Ema, R. McGehee, M. Pospelov, and A. Ray, Dark matter catalyzed baryon destruction, Phys. Rev. D \textbf{111}, 023005 (2025).
\bibitem{Berryman2024} J. M. Berryman, S. Gardner, and M. Zakeri, How macroscopic limits on neutron-star baryon loss yield microscopic limits on non-standard-model baryon decay, Phys. Rev. D \textbf{109}, 023021 (2024).
\bibitem{Reardon2024} D. J. Reardon \textit{et al.}, The neutron star mass, distance, and inclination from precision timing of the brilliant millisecond pulsar J0437$-$4715, Astrophys. J. Lett. \textbf{971}, L18 (2024).
\bibitem{Durant2012} M. Durant, O. Kargaltsev, G. G. Pavlov, P. M. Kowalski, B. Posselt, M. H. van Kerkwijk, and D. L. Kaplan, The spectrum of the recycled PSR J0437$-$4715 and its white dwarf companion, Astrophys. J. \textbf{746}, 6 (2012).
\bibitem{SINDRUM2006} W. Bertl \textit{et al.} (SINDRUM II Collaboration), A search for $\mu$-$e$ conversion in muonic gold, Eur. Phys. J. C \textbf{47}, 337 (2006).
\bibitem{Martinez2015} J. G. Martinez \textit{et al.}, Pulsar J0453+1559: A double neutron star system with a large mass asymmetry, Astrophys. J. \textbf{812}, 143 (2015).
\bibitem{MuCap2013} V. A. Andreev \textit{et al.} (MuCap Collaboration), Measurement of muon capture on the proton to 1\% precision and determination of the pseudoscalar coupling $g_P$, Phys. Rev. Lett. \textbf{110}, 012504 (2013).
\bibitem{Suzuki1987} T. Suzuki, D. F. Measday, and J. P. Roalsvig, Total nuclear capture rates for negative muons, Phys. Rev. C \textbf{35}, 2212 (1987).
\bibitem{Husain2022} W. Husain, T. F. Motta, and A. W. Thomas, Consequences of neutron decay inside neutron stars, J. Cosmol. Astropart. Phys. \textbf{10} (2022) 028.
\bibitem{Kalita2026} P. J. Kalita, S. Singh, P. Routaray, and B. Kumar, Neutron-star dark-matter limits measure the maximum-mass headroom, \href{https://arxiv.org/abs/2610.25240}{arXiv:2610.25240}.
\bibitem{AlfordHan2013} M. G. Alford, S. Han, and M. Prakash, Generic conditions for stable hybrid stars, Phys. Rev. D \textbf{88}, 083013 (2013).
\bibitem{Weissenborn2012} S. Weissenborn, D. Chatterjee, and J. Schaffner-Bielich, Hyperons and massive neutron stars: vector repulsion and SU(3) symmetry, Phys. Rev. C \textbf{85}, 065802 (2012).
\bibitem{Grinstein2019} B. Grinstein, C. Kouvaris, and N. G. Nielsen, Neutron star stability in light of the neutron decay anomaly, Phys. Rev. Lett. \textbf{123}, 091601 (2019).
\bibitem{RafieiKarkevandi2026} D. Rafiei Karkevandi, M. Shahrbaf, H. R. Moshfegh, R. Karimi, and G. Baym, Probing in-medium neutron-to-dark-fermion conversion with multimessenger neutron star observations, \href{https://arxiv.org/abs/2610.34234}{arXiv:2610.34234}.

\bibitem{BohrMottelson} A. Bohr and B. R. Mottelson, \textit{Nuclear Structure}, Vol.~I (Benjamin, New York, 1969).
\bibitem{Schwierz2007} N. Schwierz, I. Wiedenh\"over, and A. Volya, arXiv:0709.3525.
\end{thebibliography}
\end{document}